\documentstyle[times,pramana,epsf,floats]{ias}
\begin{document}

\def\lsim{\lower.5ex\hbox{$\; \buildrel < \over \sim \;$}}
\def\gsim{\lower.5ex\hbox{$\; \buildrel > \over \sim \;$}}

\mark{{Neutrino Asymmetry around black holes}{B. Mukhopadhyay and P. Singh}}
\title{Neutrino-Antineutrino Asymmetry around Rotating Black Holes}

\author{Banibrata Mukhopadhyay and Parampreet Singh}
\address{Inter-University Centre for Astronomy and Astrophysics,
Post Bag 4, Ganeshkhind, Pune-411007, India}
\abstract{
Propagation of fermion in curved space-time generates gravitational 
interaction due to the coupling between spin of the fermion and space-time 
curvature. This gravitational interaction, which is an axial-vector appears 
as CPT violating term in the Lagrangian. It is seen that this space-time 
interaction can generate neutrino asymmetry in Universe. If the back-ground 
metric is spherically asymmetric, say, of a rotating black hole, this 
interaction is non-zero, thus the net difference to the number density of the
neutrino and anti-neutrino is nonzero.
}

\maketitle
\section{Introduction}

Generation of neutrino asymmetry in early Universe is an well known fact.
If the baryon and lepton numbers
are different in our Universe, then since our Universe is electrically
neutral, it is argued that lepton asymmetry manifests itself
in the form neutrino asymmetry. Large lepton asymmetries can arise in early
Universe through for e.g. Affleck-Dine  mechanism \cite{admcdonald}.
Since relic neutrino asymmetry
has  important effects and hence can be constrained by big-bang 
nucleosynthesis and power spectrum of cosmic microwave background, 
it becomes an important issue by itself to study new mechanisms which 
can generate neutrino asymmetry specially in the present epoch.
When Dirac neutrinos propagate in gravitational backgrounds, then depending
upon the form of the background metric there is always a possibility of
the origin of neutrino degeneracy. This neutrino degeneracy is additional to
the relic neutrino asymmetry from the big-bang era.
Such an effect can have wide ranging
implications for our understanding of the phenomenon in neutrino astrophysics.

When a spinning test particle propagates in the
curved spacetime, the coupling of its spin with the spin connection of
the background field produces an interaction term. This interaction
appears in a manner similar like  a test spinor propagates in an
electromagnetic field on a background of flat space \cite{m00}.
The spin connection in curved spacetime plays a similar role as
of electromagnetic four-vector potential in the flat space.
It is very interesting to note that the interaction term
would not preserve CPT if the background spacetime contribution
does not flip sign under CPT transformation.
When the spinor under consideration
is chosen as a Dirac neutrino, this interaction under CPT will give rise to
opposite sign for a left-handed (neutrino) and right-handed (anti-neutrino)
fields.

\section{Formalism}

The most general Dirac Lagrangian density can be given as
\begin{equation}
{\cal L}=\sqrt{-g}\left(i \, \bar{\psi} \, \gamma^aD_a\psi- m \, \bar{\psi}\psi\right),
\label{lag}
\end{equation}
where the covariant derivative and spin connection are defined as
\begin{eqnarray}
D_a=\left(\partial_a-\frac{i}{4}\omega_{bca}\sigma^{bc}\right), \hskip0.5cm
\omega_{bca}=e_{b\lambda}\left(\partial_a e^\lambda_c+\Gamma^\lambda_{\gamma\mu} e^\gamma_c
 e^\mu_a\right). \label{om}
 \end{eqnarray}
 where $\sigma^{bc}=\frac{i}{2}\left[\gamma^b,\gamma^c\right]$ is the
 generator of tangent space Lorentz transformation, vierbiens are defined as 
 $e^\mu_a \, e^{\nu}_b \, g_{\mu\nu}\, = \, \eta_{ab}$.
 the Latin and Greek alphabets indicate
 the flat and curved space coordinate respectively and we would work in units
 $c = \hbar = k_B = 1$ and signature as $(+ - - -)$.
 
 Expanding (\ref{lag}), we get the Dirac Lagrangian separated into free and
 (gravitational) interaction parts as
 \begin{eqnarray}
 {\cal L}=\sqrt{-g}\,\, \bar{\psi}\left[(i\gamma^a\partial_a-m)+\gamma^a\gamma^5 B_a\right]\psi
 ={\cal L}_f+{\cal L}_I,
 \label{lagf}
 \end{eqnarray}
 where
 $B^d=\epsilon^{abcd} e_{b\lambda}\left(\partial_a e^\lambda_c+\Gamma^\lambda_{\alpha\mu}
 e^\alpha_c e^\mu_a\right)$.
Clearly, ${\cal L}_I$ is an axial-vector term which is odd under CPT transformation, 
if $B_a$ does not flip its sign. According
to the standard model, neutrino (particle) has left chirality and anti-neutrino (anti-particle)
has right chirality and thus on further expansion ${\cal L}_I$ picks up different sign for
neutrino ($\psi$) and anti-neutrino ($\psi^c$) as
\begin{eqnarray}
\overline{\psi}\gamma^a\gamma^5\psi=\overline{\psi}_L\gamma^a\psi_L,\hskip0.3cm
\overline{\psi^c}\gamma^a\gamma^5\psi^c=-\overline{\psi^c}_R\gamma^a\psi^c_R.
\label{pvec}
\end{eqnarray}
Therefore, the dispersion relation becomes
\begin{eqnarray}
E_{\nu,\overline{\nu}} =  \sqrt{|{\vec p}|^2 \pm 2 \left(B_0 p^0 + B_i p^i \right) + B_a B^a - m^2}. 
 \label{dis}
 \end{eqnarray}
Finally, if the neutrinos are traveling between two extreme points, $R_i$ and $R_f$, 
their asymmetry in number density can be generated as \cite{sm03}
\begin{eqnarray}
\Delta n=\frac{g}{(2\pi)^3}\int_{R_i}^{R_f} dV \int d^3 |{\vec p}|
\left[\frac{1}{1+exp(E_{\nu}/T)}-\frac{1}{1+exp(E_{{\overline{\nu}}}/T)}\right],
\label{fn}
\end{eqnarray}
where $dV$ is the small volume element in that space.
If we consider cartesian coordinate system, for $B_0=0$, $\Delta n=0$.


Kerr metric in cartesian coordinate system can be given as
\begin{equation} 
\hskip-1.5cm
d s^2 = d t^2 - d x^2 - d y^2 - d z^2 - \frac{2 M r^3}{r^4 + a^2 z^2} \, \bigg[ d t - \frac{1}{r^2 + a
^2} \left( r (x \, dx + y \, dy) + a(x \, dy - y \, dx) \right) - \frac{z}{r} \, dz \bigg]^2
\label{kerr}
\end{equation}
where $r$ is defined through, $r^4 - r^2 (x^2 + y^2 + z^2 - a^2) - a^2 z^2 = 0$.
As this metric yields a non-vanishing space-space cross terms and hence $B_0\neq 0$ and the
asymmetry. For detail discussions, see \cite{ms03}.

\section{Result}

Let us consider a special case of Kerr geometry where $\vec{B}.\vec{p} <<B_0 p^0$, $B_a B^a <<1$
and $B_0 <<T$. Thus from (\ref{fn}), in ultra-relativistic regime 
$\Delta n\sim \overline{B_0} T^2$, where $\overline{B_0}$ indicates the integrated value of
$B_0$ over the space.

In case of accretion disk around black hole, $T\sim 10^{11}$ K $\sim 10$ MeV $ \sim 10^{-5}$ erg. To satisfy
the approximation of metric, black hole parameters have to be tuned in such a manner that 
$\overline{B_0}\lsim 10^{-6}$ erg and thus $\Delta n \lsim 10^{-16}$.

In case of Hawking radiation bath, $T\sim 10^{-7}\left(\frac{M_\odot}{M}\right)$ K. The primordial black
hole of mass, $M>10^{15}$ gm, still exist today. The temperature of those black holes is
$T>10^{11}$ K $= 10^{-5}$ erg. Thus if $\overline{B_0}=10^{-6}$, $\Delta n \gsim 10^{-16}$, depending on
the mass of the black holes.

If there are $N_i$ number of $i$-kind black holes with corresponding curvature pseudo-scalar
coupling as $\overline{B_0}_i$ and temperature $T_i$, the neutrino asymmetry for that particular
kind of black hole can be written as
\begin{equation}
\Delta n_i=10^{-10}\left(\frac{N_i}{10^6}\right)\left(\frac{\overline{B_0}_i}{10^{-6} erg}\right)
\left(\frac{T_i}{10^{-5} erg}\right)^2.
\label{neuasy}
\end{equation}
The sign of the asymmetry strictly depends on the different black hole parameters.
Thus the overall asymmetry for all kinds of black hole becomes
\begin{equation}
\Delta n=\sum_i \Delta n_i.
\label{neuasy1}
\end{equation}
If we check this asymmetry in earth, $\overline{B_0}_{earth}=10^{-40}$ erg, $T_{earth}=10^{-14}$ erg. Then
$\Delta n_{earth}=10^{-68}$ which is very small over relic asymmetry and thus
we do not see its effect. However, in a laboratory of earth, if we are able to thermalise
the neutrino, asymmetry could be raised.

\section{Summary}

Thus we can propose criteria to generate neutrino asymmetry in presence of gravitational field as:

\noindent (1) The space-time should be axially symmetric.

\noindent (2) The interaction Dirac Lagrangian must have an axial-four-vector term
which changes sign under CPT transformation.

\noindent (3) The temperature scale of the system should be large with respect to the energy
scale of the space-time curvature.

\end{document}